\documentclass[sigconf]{acmart}

\AtBeginDocument{%
  \providecommand\BibTeX{{%
    \normalfont B\kern-0.5em{\scshape i\kern-0.25em b}\kern-0.8em\TeX}}}

\usepackage{subfig}
\usepackage{float}

\begin{document}

\title{Stock price prediction using BERT and GAN}

\author{Priyank Sonkiya}


\email{17dcs009@lnmiit.ac.in}
\authornotemark[1]
\affiliation{%
  \institution{The LNM Institute of Information Technology}
  \city{Jaipur}
  \state{Rajasthan}
  \country{India}
  \postcode{302031}
}

\author{Vikas Bajpai}
\email{vikasbajpai@lnmiit.ac.in}
\affiliation{%
  \institution{The LNM Institute of Information Technology}
  \city{Jaipur}
  \country{India}}

\author{Anukriti Bansal}
\email{anukriti.bansal@lnmiit.ac.in}
\affiliation{%
  \institution{The LNM Institute of Information Technology}
  \city{Jaipur}
  \country{India}
}



\renewcommand{\shortauthors}{Priyank and Vikas, et al.}

\begin{abstract}  

The stock market has been a popular topic of interest in the recent past. The growth in the inflation rate has compelled people to invest in the stock and commodity markets and other areas rather than saving. Further, the ability of Deep Learning models to make predictions on the time series data has been proven time and again. Technical analysis on the stock market with the help of technical indicators has been the most common practice among traders and investors. One more aspect is the sentiment analysis - the emotion of the investors that shows the willingness to invest. A variety of techniques have been used by people around the globe involving basic Machine Learning and Neural Networks. Ranging from the
basic linear regression to the advanced neural networks people have experimented with all possible techniques to predict the stock market. It's evident from recent events how news and headlines affect the stock markets and cryptocurrencies. This paper proposes an ensemble of state-of-the-art methods for predicting stock prices. Firstly sentiment analysis of the news and the headlines for the company Apple Inc, listed on the NASDAQ is performed using a version of BERT, which is a pre-trained transformer model by Google for Natural Language Processing (NLP). Afterward, a Generative Adversarial Network (GAN) predicts the stock price for Apple Inc using the technical indicators, stock indexes of various countries,
some commodities, and historical prices along with the sentiment scores. Comparison is done with baseline models like - Long Short Term Memory (LSTM), Gated Recurrent Units (GRU), vanilla GAN, and Auto-Regressive Integrated Moving Average (ARIMA) model.

\end{abstract}

%

\keywords{Stock-market, Deep-Learning, Artificial-Intelligence, Sentiment-Analysis, NLP, GAN,  BERT}

\maketitle

\section{Introduction}

The stock market, because of its high volatility, is a new field for researchers, scholars, traders, investors and corporate. The amount of computational power and techniques that have been developed have made it possible to predict the market to an extent. Recurrent Neural Networks (RNN) have been proven to predict the time series data efficiently, but the problem of vanishing and exploding gradients gave rise to the LSTM \cite{arjovsky2016unitary}.
LSTM having a good number of parameters have a greater training time in learning the short and long term dependencies. This problem created the need for what is now called the Gated Recurrent Units (GRUs). These GRU have a gate less than the LSTM (only the reset and the update gate); hence the training time is comparatively less, and so is the efficiency better \cite{yamak2019comparison}.

The sentiment analysis plays a vital role in the markets regarding the investor's emotions and their belief in the market and on a specific stock. Different newspapers and article headlines were scrapped for the sentiment analysis for Apple Inc's news using finBERT which is a fine-tuned version of BERT on a financial corpus. BERT, known to achieve state of the art in NLP tasks, is further trained on the financial corpus to create the finBERT model available in the transformers library \cite{araci2019finbert}. The output from the sentiment analysis along with technical indicators like 7-day moving average (7-DMA), 21-day moving average (21-DMA), MACD (moving average convergence divergence), bollinger bands (upper and lower), EMA (exponential moving averages), log momentum, RSI (Relative Strength Index), closing indexes of stock market's of various countries is given for prediction to a Generative Adversarial Network (GAN). The effect of technical indicators has been described by Barboza et al \cite{oriani2016evaluating} which depicts how important these indicators are.  

GANs were initially put forward by Goodfellow et al \cite{goodfellow2014generative} for generating synthetic images, but the model proposed in this paper has been modified to generate sequences instead of 2-dimensional data. The GAN uses a GRU as a generator and a 1-dimensional Convolutional Neural Network (CNN) as a discriminator. As the word adversarial suggests, the generator and the discriminator play a min-max game against each other. The generator keeps on generating synthetic data, and the discriminator evaluates it to be fake or real. Usually, the generator starts generating data from noise ~ N(0,1) passed as a latent input vector. In contrast, the proposed model takes the sentiment scores as the input vector, which has proven in the early convergence of the model. The model predicts stock prices on 5-day, 15-day and 30-day horizons and is evaluated based on the RMSE (root-mean-squared-error). 

The organization of this paper is as follows: Section 2 presents the work done in past, related to stock prediction using artificial intelligence. Section 3 talks about theoretical concepts of the paper which includes the technical indicators and the models that are used for prediction and comparison. Section 4 involves the preparation of data from various sources and the feature engineering performed. Details of the proposed model is elaborated in section 5. Experimental evaluations, model training and results of price prediction are discussed in Section 5. Finally, concluding remarks and avenues of future research work are made in Section 7.

\section{Related Work}

Stock price prediction has been done with a variety of techniques ranging from empirical, numerical, statistical to machine learning. Starting with the data itself, Chen et al\cite{chen2019hybrid} and Long et al \cite{long2019deep} used historical prices only for predicting stock prices. On the other hand, Singh et al \cite{singh2017stock}, Patel et al \cite{patel2015predicting} have added various technical indicators to the historical data as well. 
Gonadliya et al \cite{gondaliya2021sentiment} have scraped data from sources like - twitter, rss feeds and news portals and then used Bag of Words (BoW), Tf-idf, N-grams to create different feature sets. They finally used different algorithms like - Naive Bayes, Decision Trees, Logistic Regression, Random Forest and Support Vectors Machines (SVM) to compare the classification accuracy of the stock market movement. Bharadwaj et al \cite{bhardwaj2015sentiment} described the need to incur sentiment analysis while predicting stock prices.

Moghaddam et al \cite{moghaddam2016stock} uniquely proposed a model that uses the day of the week as a feature for two different timesteps, four days and nine days. They have compared Multilayer Perceptron (MLP) and Recurrent Neural Networks ( RNN ) and concluded with the comparison in transfer functions OSS and tansgig. The former performed better with four-day timestep and later with nine days. Kesavan et al \cite{kesavan2020stock} extracted data from twitter and news events in addition to the financial data of INFOSYS from moneycontrol web portal. Taking google glove dictionary as a reference, they used LSTM and lexicon-based NLP techniques to calculate the polarity of sentiments. They finally compared the regression results of LSTM with the polarity of sentiments. 

Nabipour et al \cite{nabipour2020deep} used various technical indicators like n-day moving average, RSI, CCI of sectors, diversified financials, basic metals, non-metallic minerals and petroleum. They have used ANN, RNN, LSTM and TREE-BASED (AdaBoost, Random Forest, Decision Trees and Xgboost ) models. Taking root -mean squared error as the metric they have concluded that LSTM performs the best but takes a lot of time for training. This gave the insight of using GRU in the proposed model of this paper. Mustafa et al \cite{ildirar2015interaction} concluded in their research paper that the commodity prices (Soft, Hard, Energy) have no relation with the stock prices, especially in the south Asian countries.  Maqsood et al \cite{maqsood2020local} have experimented on a massive range of models like linear regression, support vector regression, neural networks, twitter sentiment analysis, the 2012 terrorist attack and Brexit. They concluded that neural networks had the best performance. 
Sousa et al \cite{sousa2019bert} used CNN and pretrained google model - BERT in their proposed paper and found with BERT performs better than CNN for sentiment analysis. Similarly, Li et al \cite{li2021applying} have also used BERT, LSTM’s with attention and SVM to analyze the investor sentiment in the stock market. BERT with attention was shown to perform better than LSTM with attention and SVM based on accuracy score. 
Kostadin et al \cite{mishev2020evaluation} compared more than 50 approaches of NLP tasks ranging from lexicon-based approaches to transformers. They concluded that  BERT model - distilled-RoBERTa achieves the state of the art results and is best fit for classification and sentiment analysis. Artha et al \cite{andriyanto2020sectoral} in their paper proposed a different than usual methodology to create a 2d image of the stock prices. The 2d images were given as input to a 2-d CNN for classification. Their proposed model works with 96 percent accuracy on stock price movement. 

Liu et al \cite{liu2020multi} used a hierarchical capsule-based network for extracting attentive features from news and tweets. They have succeeded in increasing the accuracy by 2.45 percent from the previous baseline HCAN and StockNet models. 
While Liu et al \cite{liu2020multi} used simple GRU after the embedding layer, Mousa et al \cite{mousa2021ti} proposed a similar model with bi-directional GRU ( bi-GRU) after the word embedding to analyze news and tweets. In addition to this, they have given candlestick and bollinger band image input to the final capsule. They comparison with LSTM, GRU, bi-GRU is done in which their proposed model outperformed. Gandhmal et al \cite{gandhmal2019systematic} have put forward a review paper covering almost every algorithm ranging from 2013 - 2018. They have concluded that only Neural Network-based techniques are capable of predicting the stock market efficiently.

Patel et al \cite{patel2021event} have used simple LSTM to forecast the stock prices. In the feature set, they have added an extra attribute of the polarity of news and headlines. Each word is assigned a polarity, and the average polarity for a day is calculated to be added as a feature to the LSTM.They have also compared the results of including sentiments v/s excluding them. It was found by them, that when sentiments are considered a feature, the model trains faster and performs efficiently. Bhanja et al \cite{bhanja2019deep} used a mixture of a layer of stacked LSTM and 1-d CNN for predicting stock prices and have compared with stacked LSTM and 1-d CNN. The mixture model of stacked LSTM and 1-d CNN layers performed better than the individuals.

Zhou et al \cite{zhou2018stock} used simple Generative Adversarial Networks with LSTM as generator and CNN as a discriminator. They have used 13 technical indicators as well. For different timestep and output horizons, they have compared GAN with LSTM, ANN, SVM and ARIMA models. The GAN outperforms the rest of the models. K. Zhang et al \cite{zhang2019stock} have also proposed a Generative Adversarial Network (GAN) but with LSTM as the generator and Multi-layer-perceptron (MLP) as the discriminator. Further, they have compared the results with LSTM, ANN (Artificial Neural Network) and SVR (Support Vector Regression). They have found that the generative adversarial network performs better in terms of the root mean square error (RMSE) and mean absolute error (MAE) values.
The above described models of the Generative Adversarial networks don't incorporate the sentiment analysis on the stock market data. 
From this literature survey we can observe that many researchers have found and proved that neural networks are capable of predicting the stock market. Among the Neural Networks, the Generative Adversarial Network performs the best, which was initially put forward by Goodfellow et al \cite{goodfellow2014generative}. Sentiment analysis is as vital as doing the regression analysis with the technical indicators. Hence our proposed method ensembles the two methodologies of predicting the stock market by doing sentiment analysis on the data and then giving the sentiment scores a latent vector input to a regression model as explained in the upcoming sections. 

\begin{figure*}[ht]
\centering
	\subfloat[]{\includegraphics[width=0.5\textwidth]{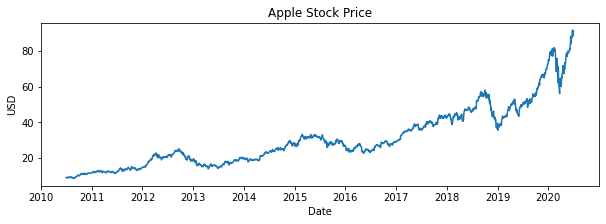}
	\label{fig:15}}\\ 
	\subfloat[]{\includegraphics[width=0.5\textwidth]{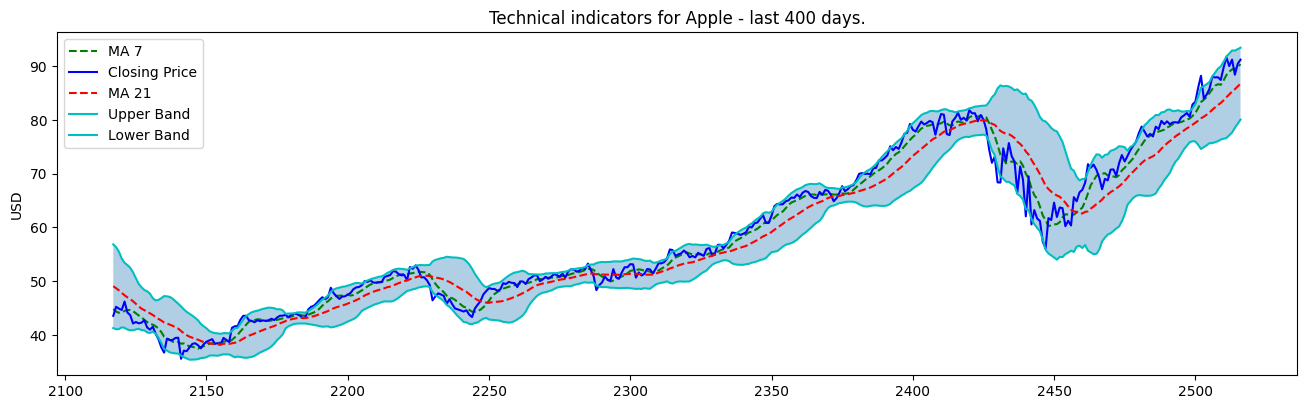}
	\label{fig:16}}\\
	\subfloat[]{\includegraphics[width=0.5\textwidth]{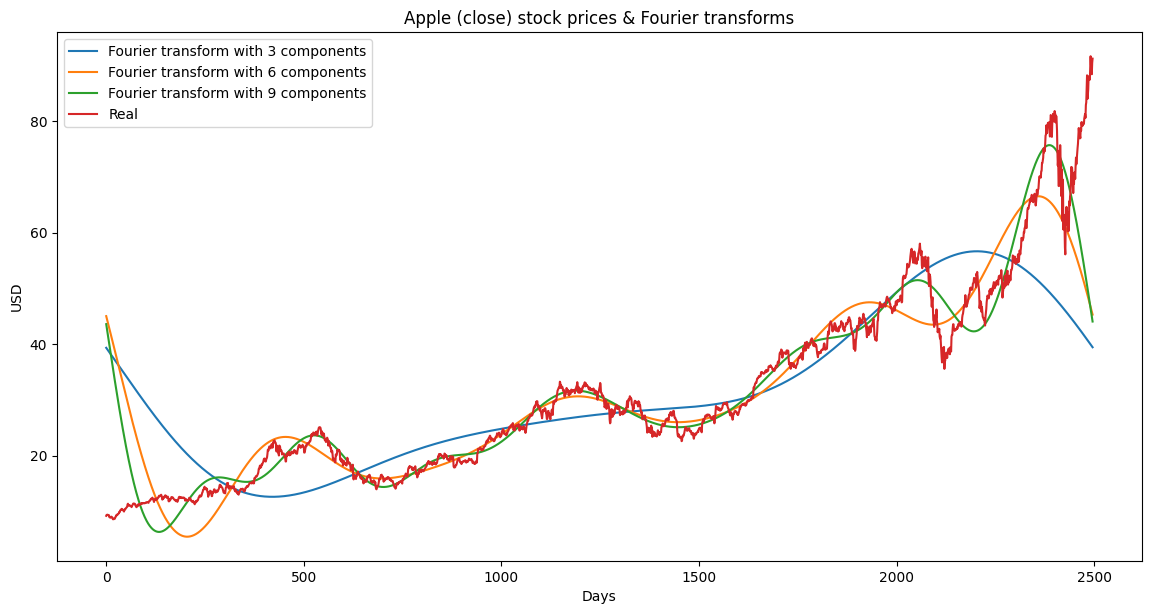}
	\label{fig:17}} 
 \caption{Apple's stock performance: (a) Apple stock price (b) Technical Indicators (c) Fourier transforms}
\label{fig:123}
\end{figure*}

\section{Theoretical Background}

The American stock market is one of the most popular and
attractive stock markets in the world. With a market
capitalization of more than 46,994,123 million dollars, the NYSE
(New York Stock Exchange) rose to 34 percent during the last
financial year. NYSE was established in 1792 as the primary
dematerialised (DMAT) institution for regulating stocks.
Predicting stock market involves the use of technical indicators
which are discussed in the next subsection. 

\subsection{Technical Indicators}
The technical
indicators are huge in number but we have discussed those
indicators which are used in our experimental setup
(Figure~\ref{fig:123}).
\begin{itemize}
\item \textbf{Moving Averages}: A simple
moving average (SMA) 7-day or 21-day is the mean of the last 7
days and 21 days respectively. It gives a good representation of
the historical prices of previous 7 and 21 days. The simple
moving average doesn’t consider the importance of recent dates.
The exponential moving average (EMA) on the other hand gives more
importance to recent dates with the help of an exponential term.
Simple moving average (SMA) = (P1+P2+P3…..PN)/N; where P1 , P2 ,
P3 are the stock prices for previous N days . 

\item \textbf{Bollinger Bands (BB)}: There are three lines that determine the
bollinger band, the simple moving average line, the upper and the
lower line. The upper and lower lines are +20 standard deviations
and -20 standard deviations away from the moving average line
respectively.  Bollinger Band = SMA  (simple moving average) ±
20SD (standard deviation) 

\item \textbf{Moving Average Convergence Divergence (MACD)}: It gives a
threshold of whether the market is overbought or oversold by
crossovers.  MACD = 26 days EMA  (exponential moving average) -
12 days EMA.  LogMomentum - Simply the logarithm of the closing
stock price of a day when aggregated gives an overall picture of
the stock trend.  
\item \textbf{RSI (Relative Strength Index)}: RSI generates
signals on bullish and bearish momentum of the market by
specifying thresholds. An asset is overbought if RSI is greater
than 70 percent and under boght if less than 30 percent.  
\item \textbf{Fourier transforms}: Breaking the closing price into trigonometric
components for maximum pattern recognition. Fourier transforms
were calculated for 3, 6, and 9 components of the closing price .
\end{itemize}

\subsection{Sentiment Analysis} 
Sentiment analysis refers to finding the intuition and emotion of a language \cite{liu2010sentiment}. This natural language processing results in the polarity of sentences classifying them as positive, negative or neutral. Sentiment analysis is highly used by business professionals and service oriented corporations that rely heavily on customer's feedback. Sentiment analysis has equal importance to evaluate a company's stock price in addition to the technical analysis.  The models to perform sentiment analysis range from lexicon based approaches to sequence-to-sequence models to transformers in the current era. The proposed model in this paper uses a modified version of BERT(Bi-Directional Encoder Representations from Transformers), the state of the art model for NLP related tasks \cite{devlin2018bert, nemes2021prediction}.

\subsection{Bi-Directional Encoder Representations from Transformers (BERT)}
BERT uses transformers\cite{vaswani2017attention} which are attentive models and are capable of establishing relationships between the words via an encoder for the input and decoder for output. The traditional NLP models take input as one word at a time whereas transformers based BERT takes the entire sentence at once to learn the pragmatic meaning hidden between the words. 


Our proposed model doesn't use BERT but a fine-tuned version of it, which is finBERT, trained on TRC2-financial data and on Financial PhraseBank \cite{malo2014good}. The fine tuned model finBERT was introduced by Araci \cite{araci2019finbert} and is trained on a huge corpus of financial data making it fit for use in financial sentiment analysis. The finBERT model is available in transformers library and can be used by creating a transformer pipeline.

\subsection{Long Short Term Memory (LSTM) }
LSTM models have improved the drawback of regular RNN’s having vanishing and exploding gradients introduced by Hochreitar et al \cite{hochreiter1997long}. LSTM’s perform exceptionally well with time series data for generating sequences. The internal structure of LSTM has 3 gates - Forget, Input and Output. Forget gate regulates the information that has to be retained or discarded. The input gate based on cell states learns the dependencies and conditions which assists in recognizing and remembering sequences. The output gate decides which information will be propagated forward as an input to the next cell.


\subsection{Gated Recurrent Units (GRU)}
GRUs are similar to LSTM but they have a gate less than LSTM. GRU’s were introduced by Kyunghyun Cho et al \cite{cho2014learning}. GRUs lack the output gate as found in the LSTM. They only consist of an update and reset gate. Just like LSTM’s, GRU is also capable of capturing sequence patterns in the data. GRU’s not having the cell state have to directly propagate information by hidden states. Since they consist of fewer parameters their training time is less and it is observed that while LSTM performs well on large datasets, GRU performs better on small datasets. 
%

\subsection{Auto Regressive Integrated Moving Average (ARIMA)}
ARIMA models were first used by George Box et al \cite{box2015time} as a mathematical approach to study changes on the time series data. ARIMA is a univariate model that tries to find patterns on its own past data via a parameter. The ARIMA model primarily focuses on making the sequence stationary since it uses lags as predictors. The predictors need to be independent of each other and not correlated. The model consists of 3 parameters (p, d, q)  where ‘p’ refers to the order of the Auto-Regressive (AR) part where it inherently means the number of lags that will be used as predictors. ‘q‘ refers to the order of the Moving-Average (MA) term, the number of lags that should go in the model. Finally ‘d’ refers to the degree of differencing that needs to be done in order to make the model stationary. 

\subsection{GAN}
 
Goodfellow et al\cite{goodfellow2014generative} developed GAN for generating images. Since then GAN’s are being modified according to the use. On the basis of regularization parameters or on the basis of type of generator and discriminator losses different versions of GAN’s have come up like - InfoGAN \cite{chen2016infogan}, time series GAN \cite{yoon2019time}. Initially the generator is fed a random noise (seed) having normal distribution and it gives some output for the discriminator to analyze. The discriminator decides as to which category the data belongs or is the distribution similar to that of the real data. In such a fashion the generator and discriminator keep fighting against each other till the point that the discriminator is confused to decide whether the data generated by the generator is fake or real. Having such functionalities the Generator is usually chosen with good regressive capabilities like LSTM’s or GRU’s whereas the discriminator is chosen with good differentiating capabilities like CNN’s. Adel et al \cite{siami2019performance} have concluded their paper with generative models performing better than the discriminative models.

\begin{figure}[!ht]
  \centering
  \includegraphics[width=0.8\linewidth]{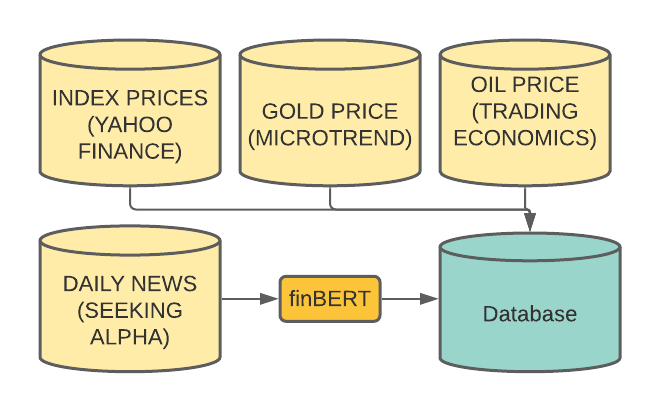}
  \caption{Data Preparation}
	\label{fig:dataPreparation}
\end{figure}

\section{Data Preparation}
Following paragraphs explain the steps involved in the data
preparation for our experiments. The same is also depicted in
Figure~\ref{fig:dataPreparation}.
The historical data for APPLE Inc. was downloaded from yahoo
finance from July 2010 till mid July 2020. The technical
indicators defined in this paper (Section 3.1) were added to the
feature set along with American stock indexes like - NYSE, NASDAQ
and S\&P 500. Apart from these, stock exchange indexes of London,
India, Tokyo, Hong Kong, Shanghai and Chicago were also used in
the feature set. Finally the historical prices of commodities
like Crude oil, Gold, US-dollar and historical prices of tech
giants like Google, Amazon and Microsoft were considered in the
data-set. The historical data of Apple from yahoo finance
contains open price, low price, high price, close price, adjusted
close price and volume of the shares traded on that day. The
final dataset contains all the above mentioned features - the
historical stock data of APPLE Inc., the technical indicators
mentioned in section 3.1, the popular stock indexes of various
countries, commodities like gold and oil. Finally in addition to
passing the sentiment vector as a latent space to the generator
the sentiment values were also used as a feature in the dataset.
The final dataset was normalized using MinMax scaler imported
from sci-kit learn. Normalization maps the range of the data from
-1 to 1 and hence reduces the complexity of the data. The
normalized value after scaling is calculated using the formulae:

\begin{equation}
x_{scaled} = \frac{x - x_{min}}{x_{max} - x_{min}}
\end{equation}

Here $x_{scaled}$ refers to the normalized value , $x_{max}$ represents the maximum attribute value and $x_{min}$ refers to the minimum attribute value.

\subsection{Data Scrapping}  
To perform sentiment analysis news and headlines were scraped from Seeking Alpha website using Beautiful Soup \cite{richardson2007beautiful} from July 2010 to July 2020. The scraped articles were cleaned and pre-processed using NLTK (Natural Language Toolkit) library \cite{bird2006nltk} and were fed to the finBERT transformer pipeline for sentiment analysis.

\subsection{Creating Timesteps}  
Time series analysis requires creating time steps for creating lookback data for which any model can learn the long term sequences of the data. Since we already have provided technical indicators with lookback data, the time series stamping was performed for 3 days only. Previous 3 day data was taken to predict the price of the 4th day data. 70 percent of the data was sent for training and the rest 30 percent for testing. The original shape of the data was (2517, 37) where 2517 is the total number of days and 37 stands for the number of features. This database after creating the time-step based look back data got converted to (2494, 3, 37) where 3 stands for the time-step parameter. After the train-test split the shape of training data was (1746, 3, 37)  and the shape of testing data was (748, 3, 37). 

\begin{figure}[htp]
  \centering
  \includegraphics[width=\linewidth]{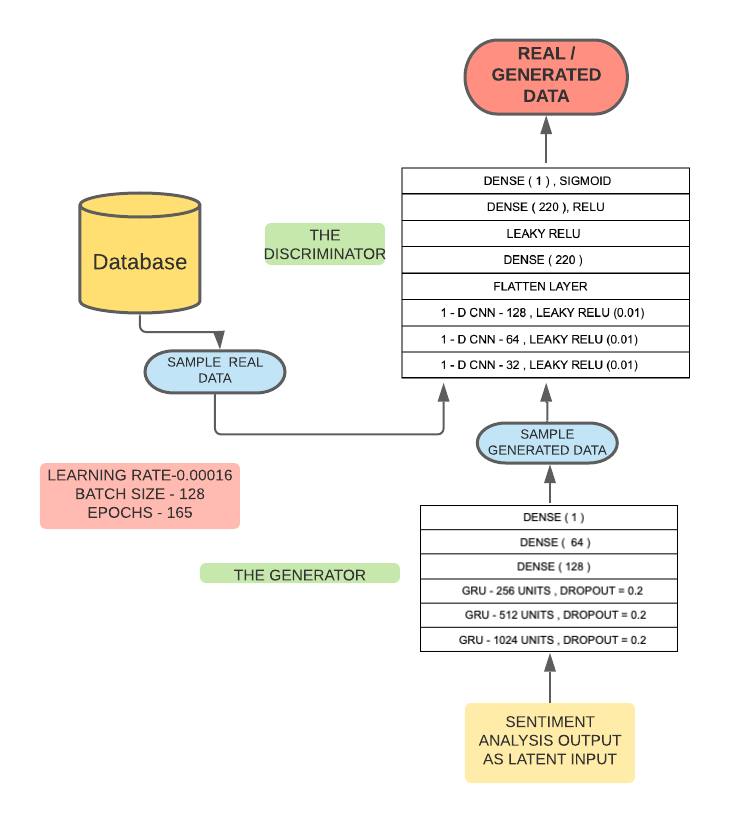}
  \caption{Proposed Model}
	\label{fig:proposedModel}
\end{figure}

\section{Proposed Model}

All the experimental programs are coded using Keras
\cite{chollet20202015} API of TensorFlow framework \cite{abadi2016tensorflow} \cite{gulli2017deep}. The
hardware setup includes Nvidia DGX-1 server with
Dual 20-core intel Xeon e5-2698 v4 2.2 Ghz processor,
supported by 512 GB DDR4 RAM and 256 GB (8x32)
GPU memory. 


The proposed model takes GRU as the generator . The
generator is represented by equation -  \begin{equation}
g_\theta: {Z\rightarrow X} \end{equation}

 where, $\theta$ stands for the weights of the Neural Network. In
 a normal GAN, Z would represent the latent input vector which is
 N (0,1). In the proposed model, the traditional Z is replaced
 with the outputs of the sentiment analysis for early
 convergence. The generator outputs  $g_\theta$(Z) which is
 represented by $P_\theta$. The final aim of the generator is to
 select  $\theta$  in such a fashion that the output of the
 generator $g_{\theta}$(Z) is estimated similarly to the original
 distribution X. In the first iteration the generator creates synthetic data from the latent space vector and propagates that to the discriminator. The discriminator classifies the data and back propagates the error back to the generator for updating the weights. So in each iteration the generator keeps on getting better in generating the data. We use GRU as the generator for our proposed
 model. The model begins with1024 units in the first layer, 512
 in the second and 256 in the third layer. To prevent over-fitting
 of the model a dropout of 20 percent was also introduced in
 every layer. In the end 3 dense layers were added each with
 units 128, 64 and 1 respectively. The generator model takes the
 input dimension, output dimensions and the feature size as
 input. The generated output is the next day’s stock prices which
 is passed on to the discriminator. 

The proposed model takes 1-dimensional
Convolutional Network as the Discriminator. The discriminator is
represented by

\begin{equation} D_w: {X \rightarrow [0,1]} \end{equation} In
	this equation parameter w represents the weights of the
	discriminator. For the real distribution $P_x$ the
	discriminator assigns 1 to the samples and for the
	distribution generated by the generator$ P_\theta$ it assigns
	0.  In the proposed GAN model, 1-d CNN was used as the
	discriminator. 32, 64 and 128 units were taken in the first
	three layers of the CNN respectively with LeakyRelu as
	activation function. Alpha value of LeakyRelu was taken to be
	0.01 in all three layers. The kernel size of 3, 5 and 5 was
	chosen in the first three layers respectively. Next a flatten
	layer was deployed to convert the data into 1-dimension
	followed by a dense layer with 220 units.  Further a layer of
	LeakyRelu was used with default alpha parameter ( 0.3 )
	followed by  another dense layer again with 220 units and
	activation function as Relu. Finally a dense layer with 1
	unit and sigmoid activation function was used in the end to
	estimate whether the data it receives belongs to the real
	data or the generated data by the generator. After classification of the data the discriminator calculates the loss between the two distributions - real and generated and back propagates to the generator as a signal to update its weights and create a better distribution which is similar to the real one. 
The loss function of a GAN is based on the KL-JS divergence. In
the training step the GAN uses cross-entropy to differentiate
between the two distributions - the real distribution and the
generated distribution. Minimizing this cross-entropy results in
minimizing the KL-JS divergence resulting in similar
distributions.

 The GAN model is represented by $(\theta,w)$. Both the individual
neural networks work against each other. The generator is trying
to minimize the difference between real and generated
distributions whereas the discriminator is trying to maximize it.
This can be represented in form of a min-max equation: 

\begin{equation} \min_\theta  \max_w E[log(D_w(X)) +
log(1-D_w(g_\theta(Z)))] \end{equation}

In equation (4) the objective of the discriminator is to
correctly differentiate between the real distribution of stock
prices v/s the generated distribution. For achieving the same,
the discriminator tries to maximize the probabilities in the
equation so that the generator output samples are classified as
generated ones $D_w(g_\theta(Z))$ and the original distribution
samples as real one $D_w(X)$.  The generator wants the
discriminator to classify the generated sample $D_w(g_\theta(Z))$
as belonging to the real distribution. Hence the generator tries
to minimize equation (4) for fooling the discriminator.  Finally,
the expectation (E) of the min-max equation makes sure that the
back propagation of error not only takes place for a sample but
for the entire data . 

Taking KL-JS divergence into consideration the min-max problem
boils down to - $\min_\theta$ JS($P_x$ || $P_\theta$)

 TRAINING - Using binary cross entropy as a differentiating
measure the weights w of the discriminator move in the positive
direction in contrast to $\theta$ of the generator which move in
the opposite direction to minimize the difference between two
distributions. For each iteration samples from the original
distribution is taken as $\{X_1,...,X_N\}$  and the corresponding
generator outputs are calculated as
$\{g_\theta(Z_1),....g_\theta(Z_N)\}$ . The same process is
repeated for the remaining iterations and the parameters are
updated according to the loss function: 

\begin{equation} w\rightarrow w + \eta  \sum^{n}_{1}\nabla_w
[log(D_w(X_i)) + log(1- D_w(g_\theta(Z_i)))] \end{equation}

\begin{equation} \theta \rightarrow \theta - \eta
\sum^{n}_{1}\nabla_\theta  [log(1- D_w(g_\theta(Z_i)))]
\end{equation}

where, $\eta$ denotes the learning rate.  The final proposed
model is presented in the Figure~\ref{fig:proposedModel}. 

\begin{table} 
	\caption{Comparison of proposed model with
	traditional approaches and the work of Zhang et al}
	\begin{tabular}{p{2.5cm}p{2.5cm}p{2.5cm}} 
	  \toprule Models & RMSE (Training) & RMSE
	  (Testing)\\ 
	  \midrule ARMIA & 10.366 & 18.2469\\ 
	  LSTM & 1.66 & 2.939\\
	  GRU & 0.70025 & 2.96\\ 
	  GAN & 0.583 & 2.369\\ 
	  GAN (Zhang et al) \cite{zhang2019stock} & Not Available & 4.1026\\
	  Proposed Model (S-GAN)  & 0.5606 & 1.827\\
  \bottomrule 
	  \label{tab:Comparison}
  \end{tabular} 
\end{table}

\begin{figure}[!ht] \centering 
	\subfloat[]{\includegraphics[scale=0.20]{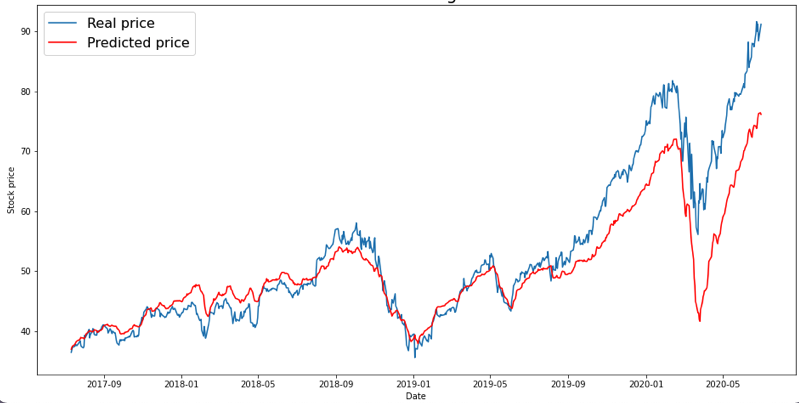}
	\label{fig:Arima}}\\
	\subfloat[]{\includegraphics[scale=0.20]{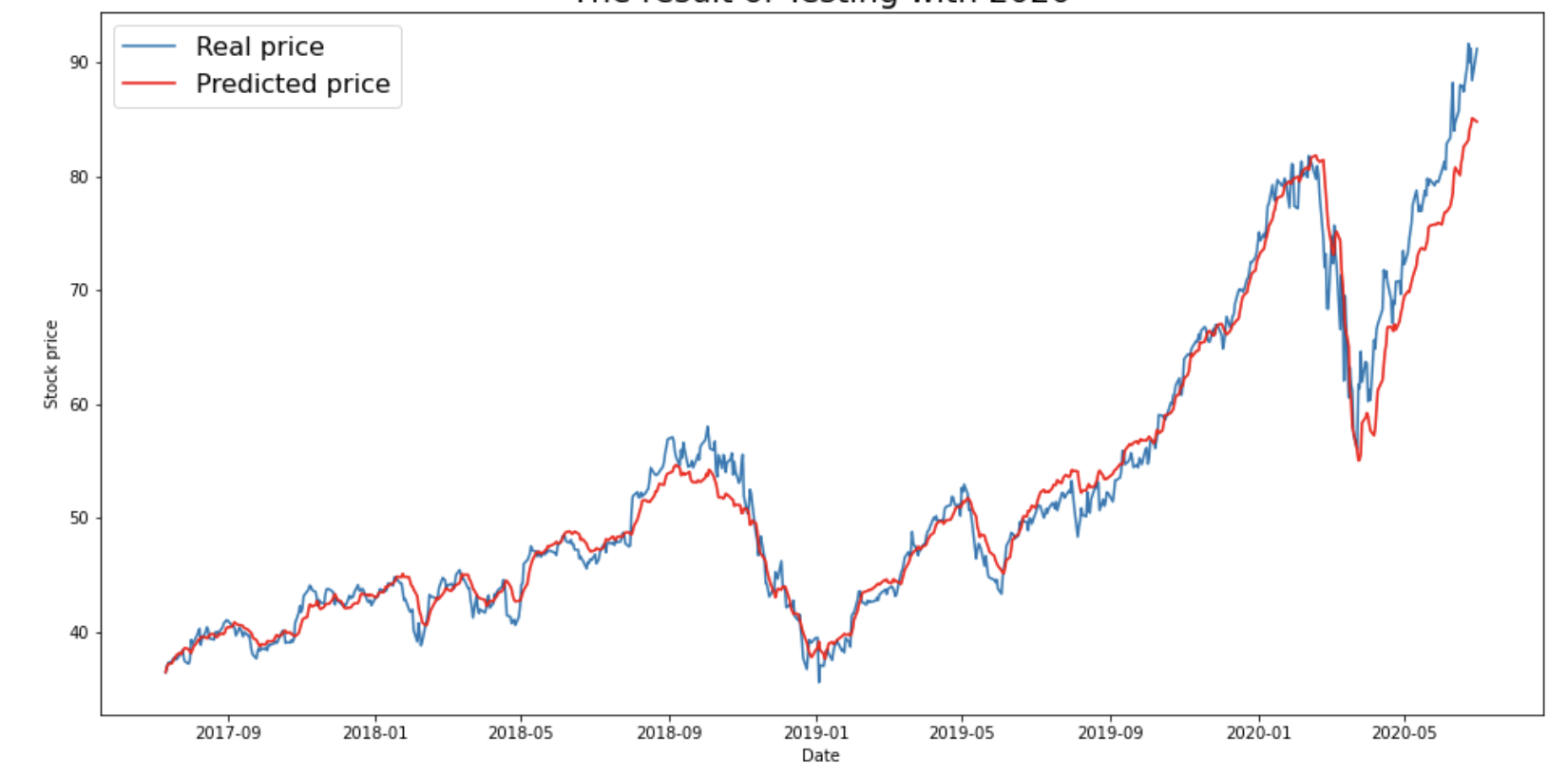}
	\label{fig:LSTM}}\\ 
	\subfloat[]{\includegraphics[scale=0.20]{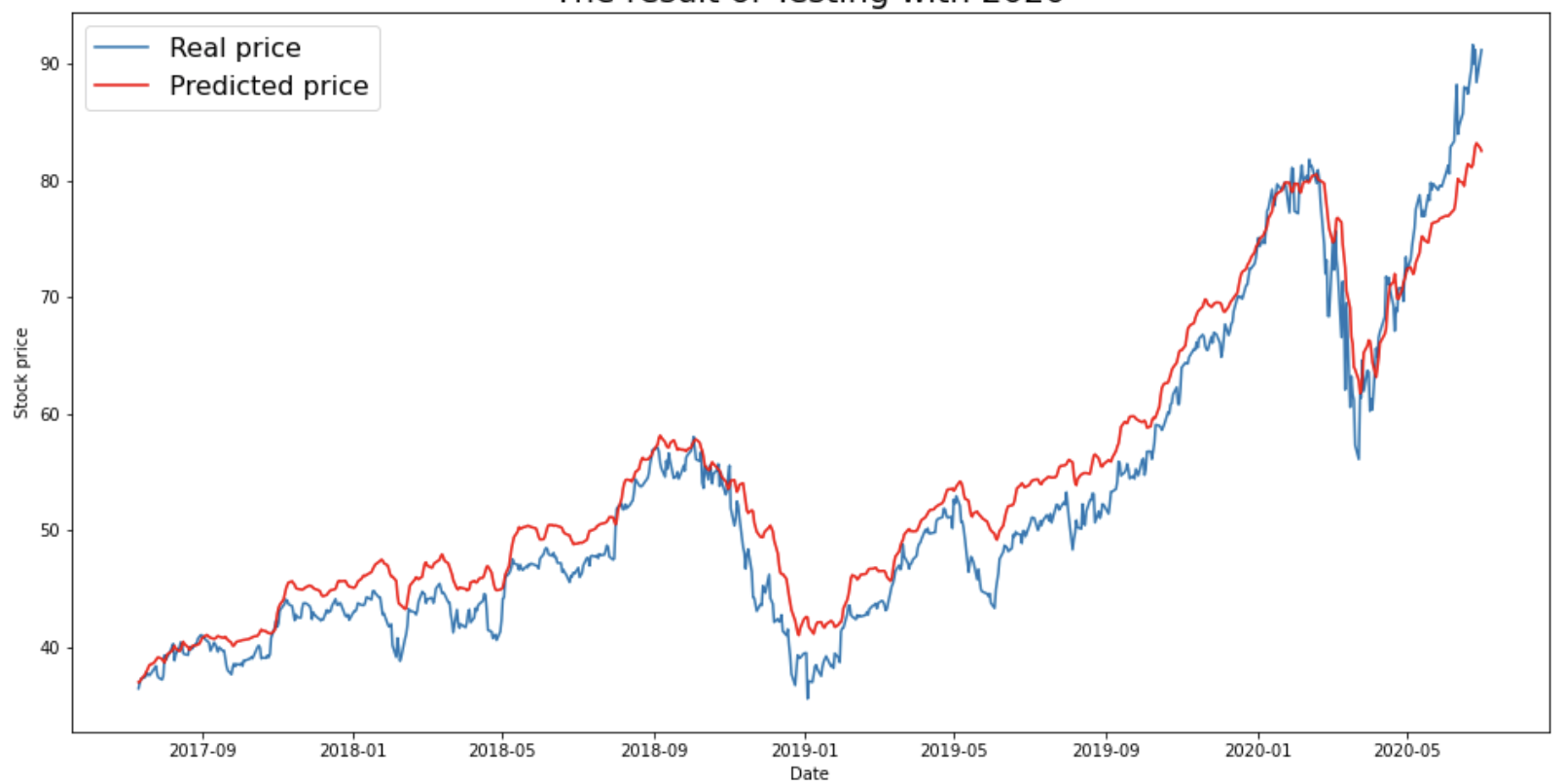}
	\label{fig:GRU}}\\
	\subfloat[]{\includegraphics[scale=0.20]{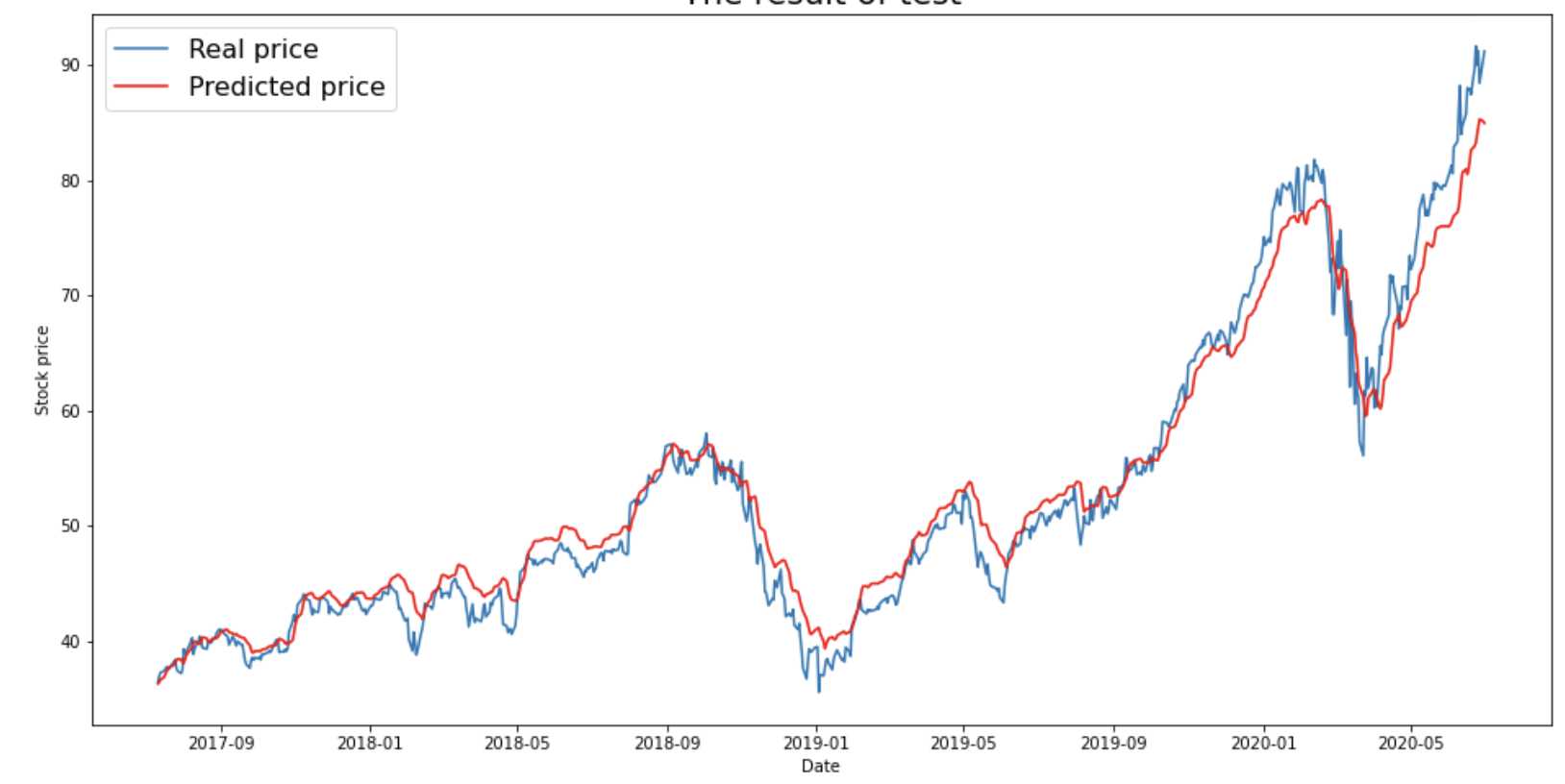}
	\label{fig:GAN}}\\
	\subfloat[]{\includegraphics[scale=0.20]{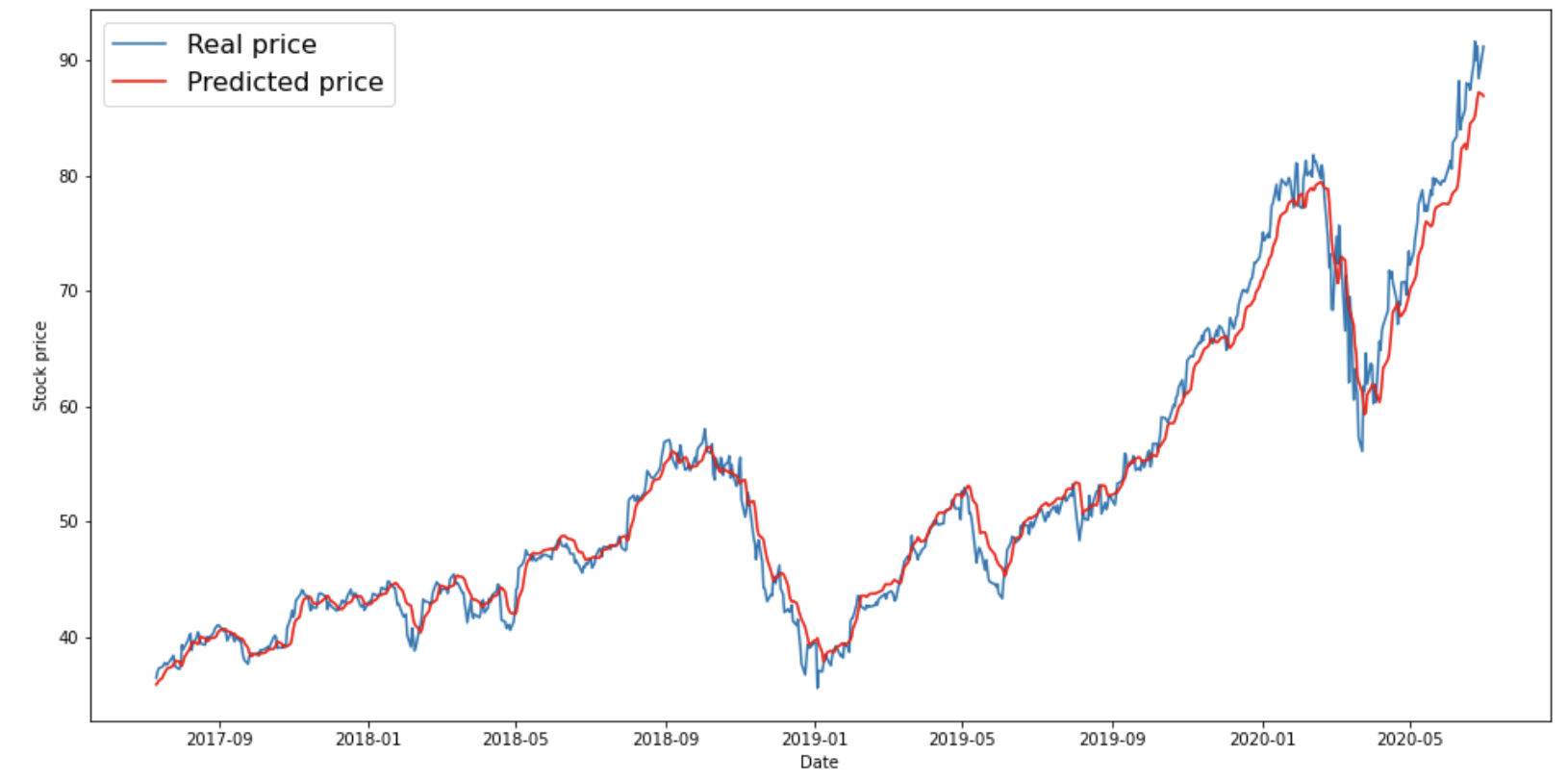}
	\label{fig:SGAN}} 	
	\caption{Comparison
	of various models and proposed S-GAN on Test Data. (a)ARIMA
	(b) LSTM (c) GRU (d) GAN (e) Proposed S-GAN }
	\label{fig:Comparison}
\end{figure}
\section{Experimental Results} Along with the proposed model
ARIMA, LSTM, GRU and plain vanilla GAN with noise as latent input
were also evaluated for comparison. For evaluating the model and
for comparison Root Mean Square Error (RMSE) metric was chosen.
Finally the models were compared on 5-day, 15 day, and 30 day
predictions. 

\begin{equation} RMSE =    \sqrt{ \frac{\sum^{n}_{1}(
Predicted_i-Actual_i)^2}{N} } \end{equation}

Here predicted refers to the predicted stock price and actual is
the real stock price. N denotes the number of data points (number
of days). 

 Auto regressive integrated moving average (ARIMA) model takes 3
 parameters p,d and q as discussed in the theoretical background.
 Being a univariate model ARIMA was run only on the closing price
 and to maintain the synchronization of data 3 time steps were
 used to generate the time sequenced data like in other models.
 Though increasing the timesteps could have resulted in better
 results for ARIMA but for comparison purposes we have considered
 only 3 timesteps. So, ARIMA (4,1,0) was used for training and
 ended up with a high RMSE of 10.366 . The testing data also as
 expected had high RMSE 18.2469. Again, this was done just for
 comparison purposes because every other model was run on 3 time
 step data and used technical indicators having historical prices
 but ARIMA being univariate only close price was considered. 
 
Namini et al \cite{siami2019performance} have proven the
efficiency of bi-directional Long Short Term Memory (LSTM) over
the uni-directional one. For comparison we have used
bidirectional LSTM with 128 cells with two dense layers having 64
and 1 unit respectively. With the loss function as simple mean
square error (MSE) the model was trained for 150 epochs, with a
batch size of 64 and learning rate of 0.0012. Adam optimizer was
used to update the weights and on the train set a RMSE of 1.66
was observed from the year 2010 to 2017. The testing RMSE was
2.939 on the data after 2017 which was used for evaluating the
model.
\begin{figure*}[!ht] \centering
	\subfloat[]{\includegraphics[width=6cm, height=3cm]{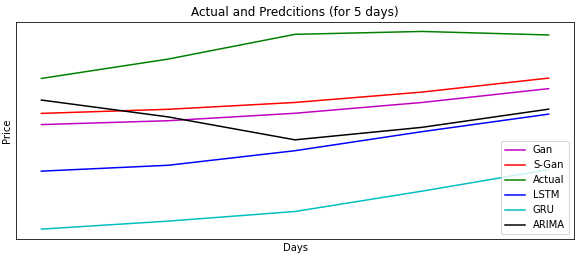}
	\label{fig:15}} 
	\subfloat[]{\includegraphics[width=6cm,	height=3cm]{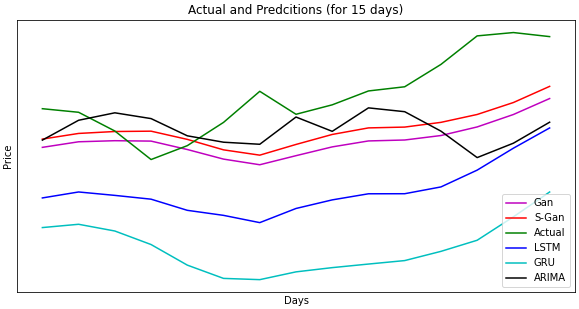} 
	\label{fig:16}}
	\subfloat[]{\includegraphics[width=6cm, height=3cm]{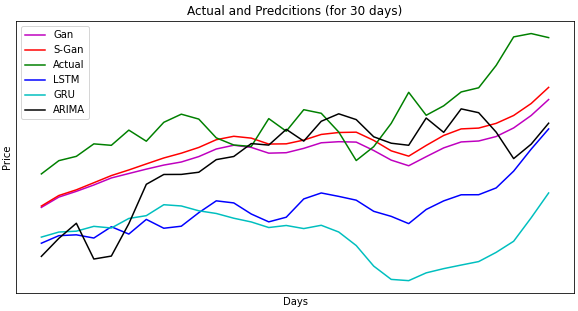}
	\label{fig:17}} 
 \caption{Performance of S-GAN and
other traditional models (a) 5 days (b) 15 days (c) 30 days}
\label{fig:ComparisonDayWise}
\end{figure*}

Gated Recurrent Units (GRU), the third model for comparison was
built with two GRU cell layers with 128 and 64 units respectively
along with two dense layers of 32 and 1 units. Mean squared error
(MSE) was used as a loss function and like LSTM, Adam optimizer
was used to update the network weights. With learning rate
0.0005, batch size as 128 and number of epochs as 50 RMSE of
0.70025 was observed on the train set. On the test data the
calculated RMSE was 2.96. As expected the GRU works similar to
LSTM with less parameters and less training time.

The plain Generative Adversarial Network (GAN) with normal
distribution as noise was trained with the same hyper-parameters
as shown in figure 3, without sentiment vector as latent noise.
With learning rate 0.00016, batch size 128 and epochs as 165 Adam
optimizer was used to update the weights. The training RMSE was
observed to be 0.583 on training data 2010-2017. The testing RMSE
calculated was 2.369 on data from 2017 till end. 

Our proposed model (named S-GAN) uses sentiment analysis output
as a latent vector and has the same hyper-parameters as the
vanilla model discussed in the above subsection for direct
comparison of results. So with learning rate as 0.00016, batch
size 128, epochs being 165, the training RMSE was observed to be
0.5606. The testing RMSE as expected turned out to be the lower
which is 1.827. As expected when given the sentiment vector as
latent input the model has proved to be efficient in two ways.
Firstly the model showed early convergence since the sentiment
vector already contains the buying and selling trends of the
market. In addition to the early convergence, the model performed
better than the vanilla model by having almost 35\% less
RMSE value and 50\% lesser RMSE than the model Zhang et al
\cite{zhang2019stock} proposed. Figure~\ref{fig:Comparison} and
Table~\ref{tab:Comparison} and shows the prediction
results of S-GAN and other traditional approaches on the Apple
Inc stock.   

Apart from the testing data in totality, predictions were
	also performed in an interval of 5,15 and 30 days for
	assessing the practicality of the model. The graphs and the
	respective RMSE values are shown in Figure~\ref{fig:ComparisonDayWise}  and Table 2.

\begin{table}[h]
	\caption{Comparison of S-GAN with other traditional
	models on 5, 15 and 30 day horizons}
	\begin{tabular}{p{1.8cm}p{1.2cm}p{1.2cm}p{1.2cm}} \toprule Models & 5 Days & 15 Days & 30
	  Days\\ \midrule ARMIA & 0.121 & 0.3658 & 0.7317\\ LSTM &
	  0.0196 & 0.0585 & 0.117\\ GRU & 0.01975 & 0.05925 & 01185\\
	  GAN & 0.0158 & 0.0474 & 0.0948\\ Proposed Model (S-GAN)  &
  0.0122 & 0.0366 & 0.0732\\ \bottomrule \end{tabular}
\end{table}

\section{Conclusion and Future Work} Our proposed model S-GAN
outperformed the traditional time series forecasting models like
GAN, GRU, LSTM and ARIMA. It's also evident from our proposed
model that in addition to the technical analysis on the stock
market, sentimental analysis also impacts the market situation by
numerically calculating the emotions and sentiment of the
investors. Our model is an advanced version of what Zhang et al
\cite{zhang2019stock} proposed that does not involve sentiment
analysis and uses LSTM as generator and MLP as discriminator.
Hence we observe better results in RMSE including sentiment
analysis as latent space while using GRU and CNN as generator and
discriminator respectively. The proposed model successfully
captures the trends in the stock market data with the historical
stock market as input data besides analyzing the sentiments. The
generator is able to confuse the discriminator whether the data
is real or is generated. Further the 5, 15 and 30 days
predictions show the applied use case of this model in daily life
stock market investments. Having all the attribute values of a
particular day, the model is able to predict the next day price.
It's also fruitful for inter-day traders who already have the
next day analysis of closing prices by the time stock market ends
for a particular day. Concluding, with the help of this model an
intuitive decision can be made the moment the stock market closes
for a day and closing prices are declared. The investors and
traders can decide based on the predicted price for the next day
whether to buy a particular stock or sell it.

The advancements in artificial intelligence have made it possible
to predict anything involving data, in such a scenario our next
goal is to focus on reward based learning and prediction. With
the help of reinforcement learning the next milestone is to
create a bot and regulate it with a reward function for getting
returns on the stock market. Further another aspect of future
work is to achieve a real time stock market prediction for
intraday trading with the help of assessing real time news and
real time stock price movements.

%

\bibliographystyle{ACM-Reference-Format}
\bibliography{acmart}

\appendix

%
%
%
%
%
%
%

\end{document}